\documentclass[aps, reprint, superscriptaddress, showpacs, longbibliography, float]{revtex4-1}
\usepackage{color, graphicx}
\usepackage{amssymb,amsmath, amsfonts, bm}
\usepackage[colorlinks,citecolor=blue, urlcolor=blue, linkcolor=blue]{hyperref}
\usepackage{bbold}
\newcommand{\angstrom}{\textup{\AA}}

\begin{document}
\title{Interacting weak topological insulators and their transition to Dirac semimetal phases}

\author{Gang Li}
\email[Correspondence and requests for materials should be addressed to: ]{gangli.phy@gmail.com}
\affiliation{\mbox{Institut f\"ur Theoretische Physik und Astrophysik,
  Universit\"at W\"urzburg, 97074 W\"urzburg, Germany}}
\affiliation{\mbox{Institute of Solid State Physics, Vienna University of Technology, A-1040 Vienna, Austria}}
\author{Werner Hanke}
\author{Giorgio Sangiovanni}
\affiliation{\mbox{Institut f\"ur Theoretische Physik und Astrophysik,
  Universit\"at W\"urzburg, 97074 W\"urzburg, Germany}}
\author{Bj\"orn Trauzettel}
\affiliation{\mbox{Institut f\"ur Theoretische Physik und Astrophysik,
  Universit\"at W\"urzburg, 97074 W\"urzburg, Germany}}
\affiliation{\mbox{Department of Physics, University of California, Berkeley, California 94720, USA}}


\begin{abstract}
Topological insulators in the presence of strong Coulomb interaction constitute novel phases of matter. Transitions between these phases can be driven by single-particle or many-body effects. On the basis of {\it ab-initio} calculations, we identify a concrete material, {\it i.e.} Ca$_{2}$PtO$_{4}$, that turns out to be a hole-doped weak topological insulator. Interestingly, the Pt-$d$ orbitals in this material are relevant for the band inversion that gives rise to the topological phase. Therefore, Coulomb interaction should be of importance in Ca$_{2}$PtO$_{4}$. To study the influence of interactions on the weak topological insulating phase, we look at a toy model corresponding to a layer-stacked 3D version of the Bernevig-Hughes-Zhang model with local interactions. For small to intermediate interaction strength, we discover novel interaction-driven topological phase transitions between the weak topological insulator and two Dirac semimetal phases. The latter correspond to gapless topological phases. For strong interactions, the system eventually becomes a Mott insulator.
\end{abstract}

\maketitle
Two classes of materials, which are in the center of focus of condensed-matter physics, are topological insulators (TIs) and strongly correlated systems.
So far, their studies have taken rather disjunct paths: the striking feature of the TIs, namely the existence of stable (Dirac) surface/edges states in a bulk insulating system, has first been predicted on the basis of ``non-interacting" band models~\cite{PhysRevLett.95.146802, PhysRevLett.95.226801, PhysRevLett.96.106802, PhysRevB.78.195424, RevModPhys.83.1057}.
In the case of the  ``Bernevig-Hughes-Zhang" (BHZ) model~\cite{Bernevig15122006}, it has then triggered the first experimental realization of a two-dimensional (2D) quantum-spin-Hall (QSH) system based on CdTe/HgTe/CdTe quantum wells~\cite{Koenig02112007}.
In this model, as well as in other semiconductor materials~\cite{PhysRevLett.103.146401, Nat.Mat.9.541, Nat.Mat.9.546} a ``band-inversion'' mechanism is crucial, where conduction and valence bands are inverted by spin-orbit coupling (SOC).
This ``non-interacting" path was soon carried over to 3D TIs~\cite{PhysRevB.76.045302,PhysRevB.75.121306,PhysRevB.79.195322,Hsien_2009,Nat.Phy.5.438,Nature_Xia}, where the search for the corresponding 2D massless Dirac states is additionally spurred by the quest for a deeper understanding of topological states of matter as well as wider spintronics applications.

Notable exceptions in this path to TIs are a recent implementation of local correlation effects into the 2D BHZ model, the Kane-Mele-Hubbard model as well as proposals for a Kondo TI state~\cite{PhysRevLett.104.106408,PhysRevB.85.045130, PhysRevLett.110.096401,PhysRevB.85.115132, Sci.Rep.4,PhysRevB.86.201407,PhysRevB.90.165136,PhysRevB.91.045122,PhysRevB.90.075140, Li:2014hk,Kim:2013gq}. 
In the study of the 2D BHZ model, some of us showed that a system with topologically trivial parameters in the absence of interactions, {\it i.e.} a band insulator, can be driven into a QSH phase by local electronic correlations~\cite{PhysRevB.87.235104}. Interestingly, this transition can even become of first-order~\cite{PhysRevLett.114.185701}.

Here, we demonstrate on the basis of density-functional theory (DFT) that a weak TI phase can be achieved in a concrete 5$d$ transition-metal compound, {\it i.e.} hole-doped Ca$_{2}$PtO$_{4}$. This discovery stimulates us to study electronic correlation effects embedded in a toy model for 3D weak TIs, which is a many-body generalization of the layered BHZ model.
We expect that the concrete material example combined with our many-body study of the layered BHZ model will provide a rich material playground for a deeper understanding of the competition between topology and interactions, in particular, in weak TIs. Somewhat surprisingly, we predict a topological phase transition from a weak TI phase to a Dirac semimetal (DSM) phase that is driven by interactions. The DSM phase is characterized by 3D bulk Dirac points that are of topological origin. The reason is that their existence (even in the non-interacting limit) is forced by the change of a 2D topological invariant as a function of one wave vector, e.g. $k_z$, if the 2D invariant is calculated for a fixed value of $k_z$.

When the most general on-site electronic correlations are included in our model via a DMFT calculation, we obtain a rich physics for small to intermediate interactions: Both the weak TI and the DSM phases are stable and the transitions between them can even be driven by electronic interaction. To the best of our knowledge, this is the first prediction for transitions between a trivial band-insulator to a DSM phase to a weak TI triggered by Coulomb interactions.

\textit{Material example:} In the first part of this Letter, we provide an example for a weak topological insulating phase in a particularlly exciting 3D system.
When examining the experimental possibility of realizing an interacting topological phase, electronic correlations and strong SOC should ideally coexist. Most of the TIs discovered in 3D are, however, semiconductors, whose nontrivial topology is generated by the strong SOC of heavy elements, such as bismuth or mercury. On the other hand, strong electronic correlations usually appear in the incompletely filled $d$- or $f$-electron shells with narrow energy bands. Thus, it would be desirable to find a nontrivial topology from a $d-p$ band inversion. Then, electronic correlations and topology could both matter. In fact, such a transition can appear in $5d$ transition-metal-oxides. In this respect, perovskite iridates~\cite{PhysRevB.85.115105, PhysRevLett.108.106401, PhysRevLett.102.256403, Balent-2010} have been theoretically predicted to be a possible host for a correlated TI phase.
Here, we propose that platinum oxides can be another promising platform for studying correlated TIs. Compared to iridates, this is rather unexplored territory. The example discussed below, {\it i.e.} the hole-doped transition metal compound Ca$_{2}$PtO$_{4}$, shows that a weak TI phase can appear in platinum oxides which has not been found experimentally in iridates so far. 

\begin{figure}[htbp]
\centering
\includegraphics[width=0.85\linewidth]{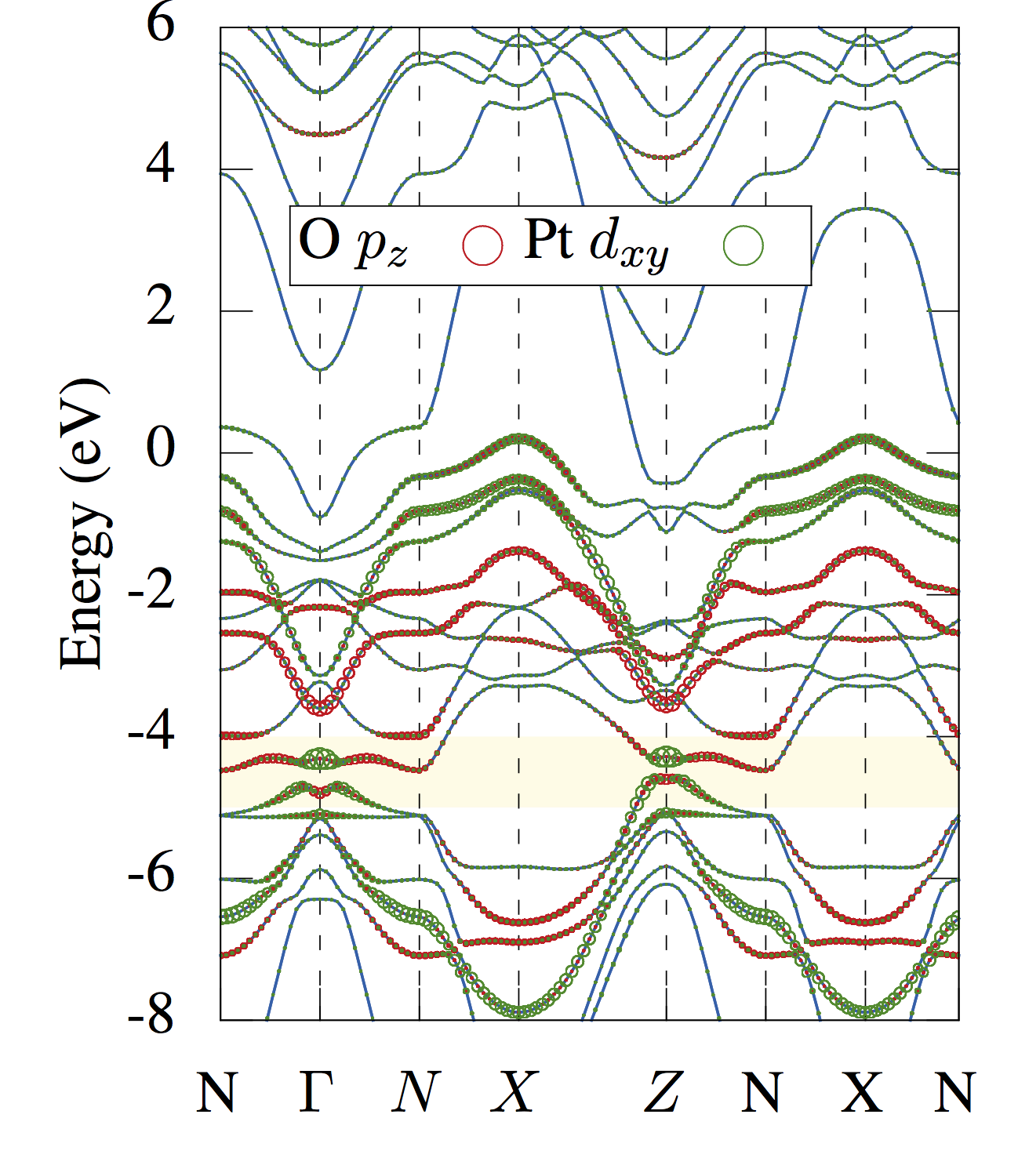}
\caption{(Color online) Electronic structure of Ca$_{2}$PtO$_{4}$ along the high-symmetry direction N($\frac{1}{2}$,0,0)-$\Gamma$(0,0,0)-N(0,$\frac{1}{2}$,0)-X($\frac{1}{2}$,$\frac{1}{2}$,0)-Z($\frac{1}{2}$,$\frac{1}{2}$,$\frac{1}{2}$)-N($\frac{1}{2}$,0,$\frac{1}{2}$)-X(0,0,$\frac{1}{2}$)-N(0,$\frac{1}{2}$,$\frac{1}{2}$). The light-yellow region shows the topological band gap which contains two inversions at $\Gamma$ and $Z$ between the bands with mainly O-$p_{z}$ (red circle) and Pt-$d_{xy}$ (green circle) characters.}
\label{Fig:Band}
\end{figure}
In Fig.~\ref{Fig:Band}, we display our electronic structure results of Ca$_{2}$PtO$_{4}$ along a high-symmetry path that connects the eight time-reversal invariant momenta (TRIMs).
The majority of the states shown are from Pt and O orbitals, whereas the Ca states only appear at above 4 eV.
DFT predicts the system to be a metal with one single band crossing the Fermi level ($E_{F}=0$) being of Pt $d_{x^{2}-y^{2}}$ character.

 \begin{figure}[htbp]
\centering
\includegraphics[width=\linewidth]{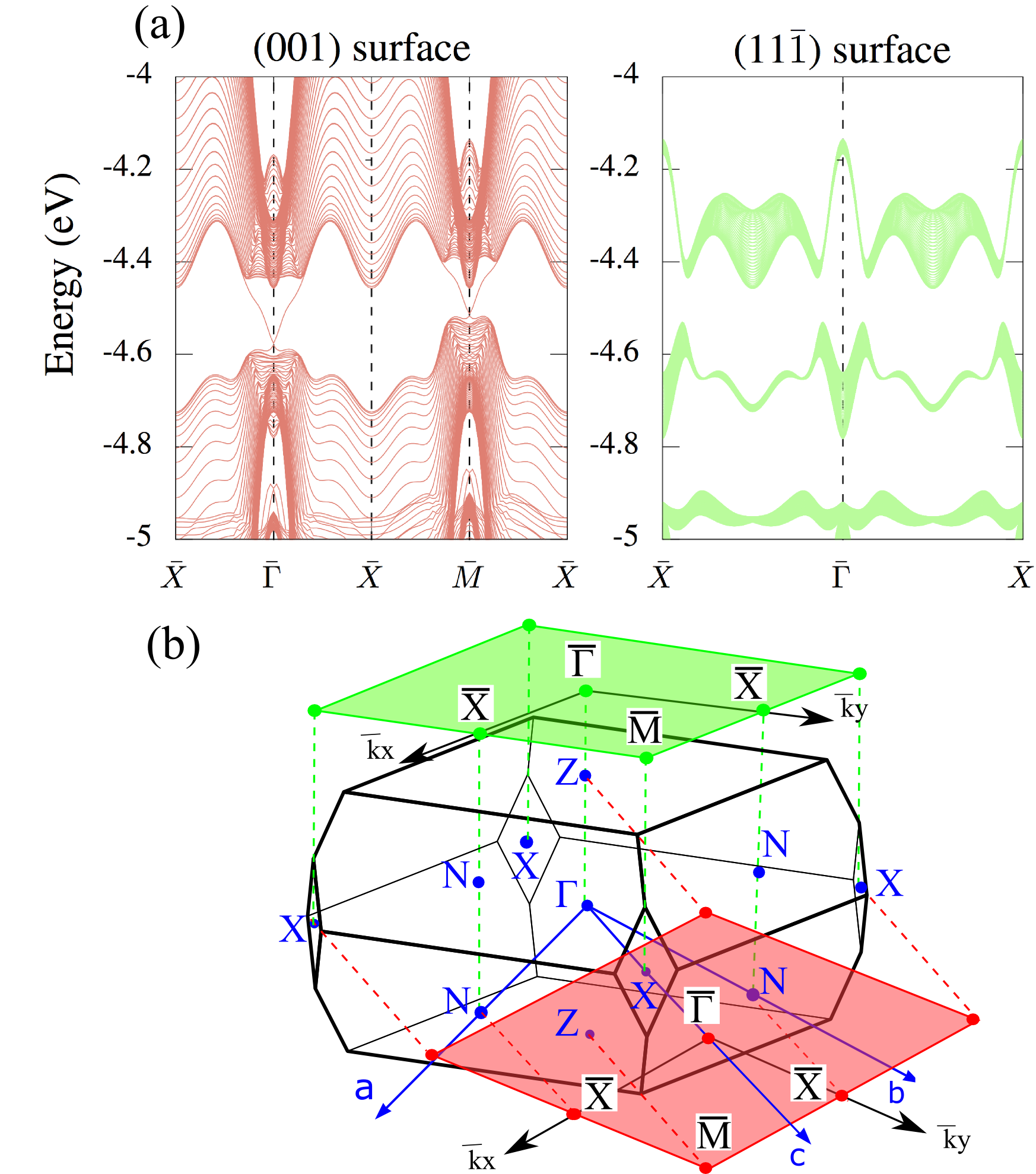}
\caption{(Color online) (a) The topological surface states show two Dirac cones at the (001) surface and no Dirac cone at the ($11\bar{1}$) surface characterizing the weak topological nature of Ca$_{2}$PtO$_{4}$. (b) The number of Dirac cones for a given surface can be determined from the relative sign of the projected pairs of the bulk TRIMs.}
\label{Fig:SBZ}
\end{figure}

In addition, we observe the striking presence of a topological gap at around -4.5 eV.
As shown in the light-yellow region in Fig.~\ref{Fig:Band}, there is a well-established energy gap of
size 100 meV, the unique feature of which is that it is due to two band inversions at $\Gamma$ and $Z$.
They are characterized by the interchange of the band characters (shown as the switch of the green and red colors occuring at the $\Gamma$ and $Z$ points of Fig.~\ref{Fig:Band} from the band below to that above the gap in the yellow region.)
Specifically, both these inversions involve O-$p_{z}$ (red circle) and Pt-$d_{xy}$ (green circle) bands. 
This topological gap is purely induced by the strong SOC of Pt.
In the Supplemental Material, we calculate the 4 topological $\mathbb{Z}_{2}$ indices $\nu_{0};(\nu_{1},\nu_{2},\nu_{3})$ as $0;(111)$, which confirms Ca$_{2}$PtO$_{4}$ to be a WTI.

A characteristic feature of a weak TI, as compared to the strong TI, is that the topological surface states (TSS) do not appear at every surface.
This behavior is demonstrated for Ca$_{2}$PtO$_{4}$ in Fig.~\ref{Fig:SBZ}.
Two surfaces, {\it i.e.} $(001)$ and $(11\bar{1})$, are considered and shown in red and green colors, respectively. Evidently, two Dirac cones appear at the (001) surface while there is no TSS found at the $(11\bar{1})$ surface.
The analysis of the TSS is slightly different from the computation of the four ${\mathbb{Z}}_{2}$ topological invariants and is presented, together with details of the slab calculations in the Supplemental Material.

The predicted topological gap lies deep in the valence band. Thus, it will be a challenge in experiments to bring the Fermi level down to the topological gap region.
As discussed before, the Ca states basically do not contribute to the band inversion, thus, it is possible to substitute Ca with other elements with less valence electrons to effectively hole-dope the system.
Potassium substitution may not be sufficient to fully bring the Fermi level down to -4.5 eV. However, a subsequent electric gating and/or a direct angle-resolved photoemission spectroscopy (ARPES) detection may then become feasible.

\textit{3D topological model:}
Our concrete example Ca$_{2}$PtO$_{4}$ seems to provide a rich playground for finding new 3D topological insulators. This statement holds provided that the obviously required inclusion of strong electronic correlations in this $5d$-material does not destroy the topological phases identified in the ``non-interacting", {\it i.e.} DFT + SOC calculations.
This problem will be analyzed next in a 3D topological toy model that resembles our material example in the vicinity of the band inversion points. Differently from the strong TI, a weak TI can be adiabatically connected to stacked layers of 2D TIs~\cite{PhysRevLett.98.106803}.
This feature motivates us to construct, in 3D, a half-filled weak TI model by coupling layers of 2D topological systems, {\it i.e.}  2D BHZ models, with a nearest-neighbor hopping $t_{z}$ along the stacking ($z$) direction. Subsequently, we study the interaction-driven topological phase transitions of this model.

The ``non-interacting" Hamiltonian on a cubic lattice is hence given by
\begin{eqnarray}\label{weak TI}
H(\mathbf{k}) &=& [M-(\cos k_{x} + \cos k_{y}  + t_{z}\cos k_{z})]\sigma_{z}\otimes I \nonumber\\
&& + \lambda\sin k_{x}\cdot \sigma_{x}\otimes S^{z} + \lambda\sin k_{y}\cdot\sigma_{y}\otimes I\;,
\end{eqnarray}
where $M$ is the mass parameter, $\lambda$ denotes the SOC strength.
$0<t_{z}<1$ is the hopping amplitude along the stacked direction (z) perpendicular to the in-plane (xy) hopping.
When $t_{z}$ is taken as zero, Eq.~(\ref{weak TI}) reduces to the 2D BHZ model.
This model is different from those traditionally introduced for the 3D strong TIs~\cite{nphys1270, PhysRevB.82.045122,PhysRevB.81.041307} in that it is still block-diagonal in spin. 
Note that the Pauli matrices $\sigma_{x,y,z}$ denote an orbital space degree of freedom and the Pauli matrices $S^{x,y,z}$ refers to the physical spin.

Instead of restricting our discussion to any specific choice of the tuning parameters (which will be done in our comparison with Ca$_{2}$PtO$_{4}$ in the Supplemental Material), we study first the Hamiltonian in Eq.~(\ref{weak TI}) from a more general perspective and explore all possible phases that a topological system, described by Eq.~(\ref{weak TI}), can display.
In the second step, we then want to understand how the topological transitions between these phases are influenced by electronic correlations.

\begin{figure}[htbp]
\centering
\includegraphics[width=\linewidth]{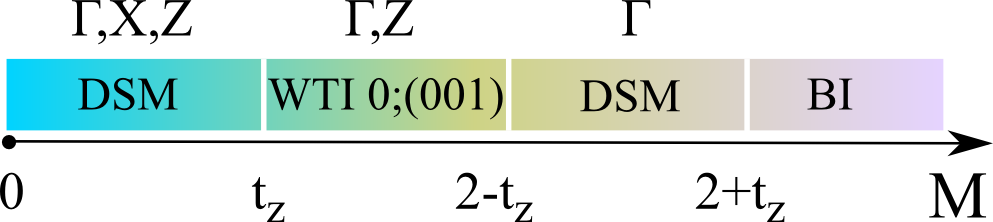}
\caption{(Color online) Depending on the strength of $M$, the 3D topological model in Eq.~(\ref{weak TI}) displays four topologically distinct phases, characterized by a different number of band-inversions as well as the presence (WTI and BI) or the absence (DSM) of a bulk gap. }
\label{Fig:TB-topology}
\end{figure}

The Hamiltonian (\ref{weak TI}) is time-reversal invariant and at the time-reversal invariant momenta (TRIM) $\mathbf{k}=\Gamma_{i}$, it is simplified to
\begin{equation}\label{TB-Z2}
H(\mathbf{k}=\Gamma_{i}) = [M-(\cos k_{x} + \cos k_{y}  + t_{z}\cos k_{z})]\sigma_{z}\otimes I \;.
\end{equation}
Thus,  $H$ commutes with the parity operator $\hat{P}=\sigma_{z}\otimes I$, implying that the eigenstates of Eq.~(\ref{TB-Z2}) are also the eigenstates of the parity operator, from which the ${\mathbb{Z}}_{2}$ invariants can be easily calculated~\cite{PhysRevB.76.045302} (see also the Supplemental Material).
As a function of the mass parameter $M$ (for a fixed value of interlayer coupling $t_{z}$), we find four topologically distinct phases, characterized by a different number of band inversions.
These phases are shown in Fig.~\ref{Fig:TB-topology}. 
We identify a weak TI phase, two DSM phases, as well as a trivial band-insulating (BI) phase. Notably, also the gapless DSM phases are topologically nontrivial~\cite{ncomms5898}. Interestingly, in addition to the linear bulk band crossing there are also band inversions at $\Gamma$ or at $\Gamma$, $X, Z$. In these phases, the definition of the 3D topological invariants that we used to characterize the weak TI phase is not valid any more (because of the gapless bulk). However, a 2D topological invariant can be calculated for any fixed $k_{z}$ plane, as long as a gap remains open in the 2D case. Additionally, a mirror Chern number can be defined to characterize the nontrivial topology of these phases~\cite{ncomms5898}.
More specifically, the 3D DSM has its topological origin in the 2D ${\mathbb{Z}}_{2}$ invariants (for a fixed value of $k_{z}$) of the corresponding 2D BHZ model. Two 2D topological invariants can then be defined at a $k_{z}> k_{z}^{c}$ and $k_{z} < k_{z}^{c}$, where at $k_{z}^{c}$ the bulk bands cross. These two 2D topological invariants differ which is the reason why the bulk gap has to close at $k_{z}^{c}$.

For the "non-interacting" 3D layered BHZ model, the topological phase transition is driven by the value of $M$. In what follows, we want to show that the same transitions can be triggered by electronic correlations as well.
To this aim, we add the most general local interaction term to Eq.~(\ref{weak TI}), {\it i.e.}
\begin{eqnarray}\label{H_u}
H_{U} &=& U\sum_{i,\alpha}n_{i\alpha\uparrow}n_{i\alpha\downarrow,\sigma}+
U^{\prime}\sum_{i,\alpha<\beta}n_{i\alpha\sigma}n_{i\beta\bar{\sigma}}\nonumber\\
&&+(U^{\prime}-J)\sum_{i,\alpha<\beta,\sigma}n_{i\alpha\sigma}n_{i\beta\sigma} \nonumber\\
&& +J\sum_{i,\alpha\ne\beta}(c_{i\alpha\uparrow}^{\dagger}c_{i\beta\uparrow}^{}c_{i\beta\downarrow}^{\dagger}c_{i\alpha\downarrow}^{} +
c_{i\alpha\uparrow}^{\dagger}c_{i\alpha\downarrow}^{\dagger}c_{i\beta\downarrow}^{}c_{i\beta\uparrow}^{})\;.
\end{eqnarray}
Here, $i$ is the site index of a 3D cubic lattice, $\alpha, \beta$ span the orbital basis of this Hamiltonian.
A rotational invariant form of $H_{U}$ is chosen by us, {\it i.e.} $U^{\prime}=U-2J$
and $J$ is fixed to a rather large but not untypical value for transition-metals, {\it i.e.} $U/4$.

Evidently, Eqs.~(\ref{weak TI}) and (\ref{H_u}) comprise the interacting many-body Hamiltonian for a TI that can demonstrate various topological phases as shown in Fig.~\ref{Fig:TB-topology} for $U=0$.
In addition to this non-interacting case, there is another limit where an explicit solution of the full Hamiltonian is known. When $U\rightarrow\infty$, the paramagnetic ground state will be a Mott insulator, which destroys all nontrivial topological phases present at weak interactions.

\begin{figure}[htbp]
\centering
\includegraphics[width=\linewidth]{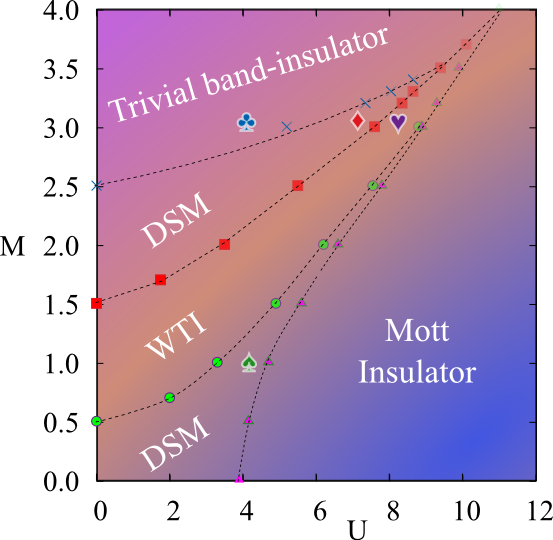}
\caption{(Color online) The interacting phase diagram of the 3D topological model based on Eqs.~(\ref{weak TI}) and (\ref{H_u}). Five different phases are identified. At sufficiently large $M$ and $U$, the conventional band insulator and the Mott insulator phases dominate. In between, the competition between Coulomb interaction and topological order gives rise to interesting phases such as a region of a weak TI phase and two regions of DSM phases. The topological surface states (TSS) with parameters given by the corresponding colored symbols can be found in the Supplemental Material. As we carefully explain there, the phase boundaries are subject to an approximation for large values of $U$.}
\label{Fig:PhaseDiagram}
\end{figure}
Before discussing the results in Fig.~\ref{Fig:PhaseDiagram}, we briefly describe the influence of electronic correlations on TIs and explain how the transition boundaries in Fig.~\ref{Fig:PhaseDiagram} have been obtained. 
Similarly to their non-interacting counterparts, the topology of correlated TIs can be defined by means of an ``effective Hamiltonian'' $h_{\text{eff}}(k)=-G^{-1}(k, \omega\rightarrow0)$~ \cite{PhysRevLett.105.256803,PhysRevB.85.165126,PhysRevX.2.031008}, which, in the DMFT language, reads $h_{\text{eff}}(k)=H(\mathbf{k})+\hat{\Sigma}(\omega\rightarrow0)$.   
For any nonzero value of $M$, the imaginary part of the self-energy always extrapolates to zero for $\omega\rightarrow0$.
The influence of correlations on the electronic structure is therefore mainly encoded in $\Re \hat{\Sigma}(\omega\rightarrow0)$.  
Thus, the transitions between the BI, the DSM and the WTI phases in Fig.~\ref{Fig:PhaseDiagram} are determined simply from an inspection of the topology of $H(\mathbf{k})+\Re \hat{\Sigma}(\omega\rightarrow0)$. 
In the Mott phase, we have however used the even more clearcut condition $n_{\alpha}=n_{\beta}=1$~\cite{PhysRevB.86.201407}. More details about the accuracy of the transition lines in the large-$U$ part of the phase diagram can be found in the Supplemental Material.

The first important observation about Fig.~\ref{Fig:PhaseDiagram} is that the topological nontrivial phases extend to a finite value of $U$. Thus, the topology of such a 3D topological system is stable against a ``certain amount'' of electronic correlations, which may pave the way for the experimental realization of correlated TIs. With the increase of $U$, each topologically nontrivial phase extends towards a larger value of $M$. This behavior can be understood because the Coulomb interaction modifies the mass parameter $M$ via  $\Re \hat{\Sigma}(\omega\rightarrow0)$.
In particular, as an effect of the nonzero value chosen for the Hund's coupling $J$, a larger value of $M$ is required to maintain the phase.

The second important observation is that, for a fixed value of $M$, the Coulomb interaction $U$ may generate transitions between different topological phases. This is again due to the inter-orbital interaction which favors single orbital occupations.
The colored spade, diamond and heart symbols in Fig.~\ref{Fig:PhaseDiagram} correspond to three topologically distinct phases with the same $M$, but at different $U$ values.
Among the phase transitions driven by the interaction $U$, the first one is of particular interest, {\it i.e.} it is the transition from a trivial band insulator to a DSM and to a weak TI, see Fig.~\ref{Fig:TB-U}(a), (b) and (c).
This observation implies that the Coulomb interaction does not always destroy but sometimes also induces topological phases.

In summary, we have investigated the stability of 3D topological phases in the presence of electron-electron interactions.
Our study is motivated by the discovery that, on the basis of a DFT band structure calculation, the hole-doped compound Ca$_{2}$PtO$_{4}$ is a weak TI with the band inversions induced by the Pt-$d_{xy}$ and O$-p_{z}$ orbitals.
This finding in principle opens up a new direction in the search for 3D interacting TIs, provided that the expected local correlations in this 5$d$ transition-metal do not destabilize the topological nature.
To verify this, we have modelled the DFT topological band structure as the underlying ``non-interacting'' band structure of a 3D generalization of the BHZ Hamiltonian, which has then been augmented by the most general on-site two-orbital electron-electron interaction.
For small to intermediate interactions, topological phases as well as transitions between them are induced by the interactions, in particular, a novel DSM to weak TI transition. For very strong interactions, the system enters the expected ``trivial'' Mott-insulating phase.

We thank A.~Amaricci, J.~C.~Budich, A.~Fleszar and K.~Haule for fruitful discussions. GL and WH acknowledge financial support by the DFG Grant No. Ha 1537/23-1 within the Forschergruppe FOR~1162, the SPP Grant Ha 1537/24-2, as well as computing time granted at the Leibniz Supercomputing Centre (LRZ) in Munich. 
GS has been supported by the DFG through the SFB 1170 ``ToCoTronics''. 
BT acknowledges financial support by the DFG (German-Japanese research unit ``Topotronics'', the priority program SPP 1666, and the SFB 1170 ``ToCoTronics''), the Helmholtz Foundation (VITI), and the ``Elitenetzwerk Bayern'' (ENB graduate school on ``Topological insulators'').

\bibliography{ref}


\pagebreak
\newpage
\onecolumngrid

\begin{center}
\textbf{\large{Supplemental Material}}\\
\end{center}

\section{Ca$_{2}$PtO$_{4}$, a hole-doped weak TI}
As a concrete example, we consider the K$_{2}$NiF$_{4}$-type structure of Ca$_{2}$PtO$_{4}$ with space group I4/mmm. The lattice parameters taken from an {\it ab-initio} geometry optimization~\cite{Matar201149}, are $a=3.87~\angstrom$ and $c=11.42~\angstrom$.
The density-functional theory (DFT) calculations were carried out within the full-potential linearized
augmented plane-wave (FP-LAPW) method~\cite{FP-LAPW}, implemented in
the package WIEN2k~\cite{Wien2k}.
$K_{max}R_{MT}=9.0$ and a $10\times10\times10$ k-mesh were used for
the ground-state  calculations.
R$_{MT}$ represents the smallest muffin-tin radius and
$K_{max}$ is the maximum size of reciprocal-lattice vectors. The spin-orbit
coupling is included by a second variational procedure. The
generalized gradient approximation (GGA) potential~\cite{perdew1996,
  perdew1997} is used in all calculations..
The surface electronic structures are further calculated using the
maximally localized Wannier functions (MLWFs)~\cite{Mostofi2008685},
employing the WIEN2WANNIER~\cite{wien2wannier90} interface.
The MLWFs are constructed in a non-self-consistent calculation with an
$8\times8\times8$ k-mesh.

\begin{figure}[htbp]
\centering
\includegraphics[width=0.6\linewidth]{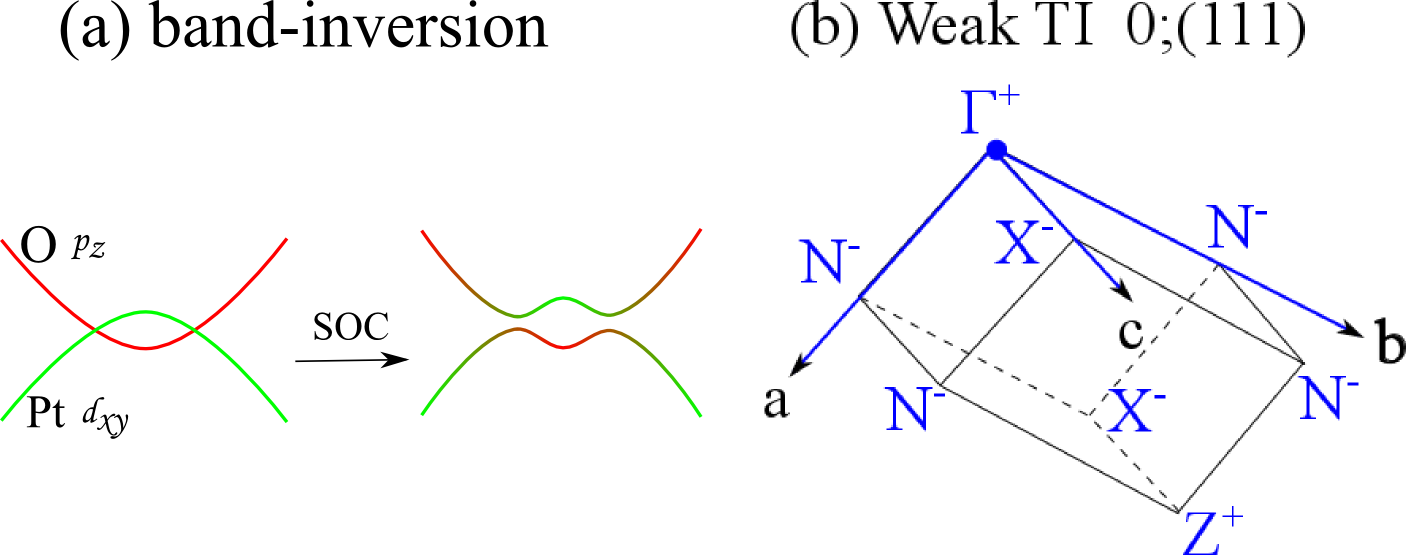}
\caption{(Color online) (a). The presence of spin-orbital coupling induces a band-inversion between O-$p_{z}$ (red line) and Pt-$d_{xy}$ (green line) bands, which characterises the nontrivial topology of this system. (b) A topological invariant can be defined for the topological gap shown in the light-yellow region of Fig.~\ref{Fig:Band}. It is found, that this gap possesses a weak-topological nature. The +/- sign associated with each high-symmetry point is the parity products for the bands below the topological gap at this TRIM. In the coordinate system defined in Fig.~\ref{Fig:SBZ}, the topological invariant is given as $0;(111)$. }
\label{Fig:Inversion}
\end{figure}
As shown in Fig.~\ref{Fig:Inversion}(a), without SOC,the O-$p_{z}$ (red line) and Pt-$d_{xy}$ (green line) bands overlap at $\Gamma$ and $Z$. In the presence of SOC, the hybridization of these two bands opens an energy gap, which leads to the inversion of the two bands at those two TRIMs.
The appearance of an even number of band-inversions in a 3D system can give rise to a weak TI~\cite{PhysRevLett.98.106803}.
Thus, Ca$_{2}$PtO$_{4}$ can become a weak TI given that the Fermi level is shifted down into the topological gap around -4.5 eV.
To more accurately verify the topological nature of this material system at the DFT level, we calculate the four topological invariants $\nu_{0};(\nu_{1},\nu_{2},\nu_{3})$ through the examination of the parity products ($\delta_{i}$) of the occupied states at all eight TRIMs (see Eqs.~(\ref{Z2-strong}), (\ref{Z2-define}), (\ref{Z2-weak})).
As there is a well-defined energy gap for the band-inversions, the separation of the occupied and unoccupied bands at this energy level certainly justifies the definition of the topological invariant.
Ca$_{2}$PtO$_{4}$ respects inversion symmetry. Thus, we can simply identify the parities of all occupied bands below this topological gap to calculate the topological invariant.
The products of the parities for all occupied bands below the topological gap ({\it i.e.} $\delta_{i}$ in Eq.~(\ref{Z2-define}))are marked as +/- sign in Fig.~\ref{Fig:Inversion}(b) for each TRIM.
The coordinate system of the BZ is chosen to be the same as that in Fig.~\ref{Fig:SBZ}(b).
The four ${\mathbb{Z}}_{2}$ indices $\nu_{0};(\nu_{1},\nu_{2},\nu_{3})$ are then easily calculated as $0;(111)$.
The strong topological index $\nu_{0}=0$, while the other three weak topological indices are nonzero which, thus, confirms Ca$_{2}$PtO$_{4}$ to be a weak TI.

Next, we summarize the topological surface states (TSS) calculation of Ca$_{2}$PtO$_{4}$, as displayed in Fig.~\ref{Fig:SBZ}.
Taking the (001) surface as an example, the surface Brillouin zone (SBZ) has four TRIMs $\Lambda_{a}$, {\it i.e.} 1$\overline{\Gamma}$, 2$\overline{X}$, 1$\overline{M}$.
They are the projections of pairs of bulk momenta $\Gamma_{a1}$, $\Gamma_{a2}$, that differ by $\mathbf{G_{c}}/2$, onto the (001) plane.
As a result, the relative sign at the SBZ TRIMs $\pi_{a}$ are given as the product of the corresponding $M_{a}$, {\it i.e.} $\pi_{a} = \delta_{a1}\delta_{a2}$.
For example, $\overline{\Gamma}$ is the projection of the bulk TRIMs $\Gamma$ and $X$ on the (001) surface (see Fig.~\ref{Fig:SBZ}(b)), $\pi_{\overline{\Gamma}}$ is then computed as $\delta_{\Gamma}\delta_{X}=-1$.
Similarly, one can derive the relative signs at $\pi_{\overline{M}}=-1$ and $\pi_{\overline{X}}=1$.
Thus, there will be two Dirac cones at $\overline{\Gamma}$ and $\overline{M}$, while there is no TSS at $\overline{X}$.
In Fig.~\ref{Fig:SBZ}(a), the TSS are displayed from a slab calculation of 8 nm thickness in the given direction.
In practice, this is achieved by projecting the DFT Bloch bands to the Pt-d and O-p orbitals, which gives rise to a tight-binding model with only the target orbitals as a basis.
Thus, compared to the full DFT parameter space, this basis set is much smaller and, thus, it becomes feasible for a larger slab calculation.
In a similar fashion, at the ($11\bar{1}$) surface, one can understand that the momenta 2$N$ project onto $\overline{X}$, 2$X$ onto $\overline{M}$, $\Gamma$ and $Z$ onto $\overline{\Gamma}$.
Thus, at all four surface TRIMs, the relative signs $\pi_{a}$ are positive. Consequently, there is no TSS at this surface, see Fig.~\ref{Fig:SBZ}(a) for comparison.

\section{A 3D topological model}

\begin{figure}[htbp]
\centering
\includegraphics[width=0.5\linewidth]{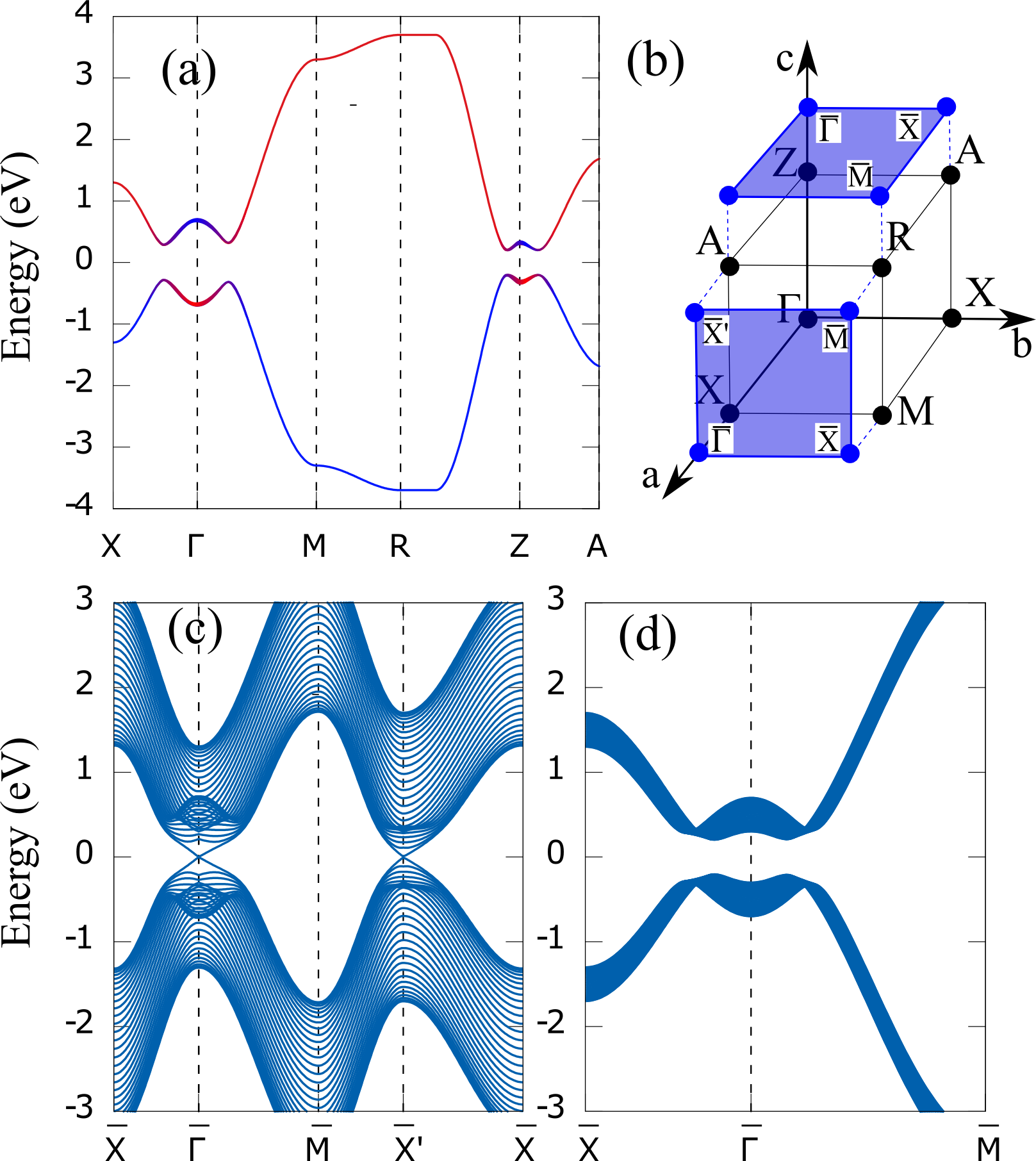}
\caption{(Color online) (a) Bulk electronic structure of the weak TI model defined in Eq.(\ref{weak TI}) along a high-symmetry path of a cubic lattice with the BZ plotted is in (b).
Two band-inversions (presented as the interchange of color) at $\Gamma$ and $Z$ give rise to topologically distinct surfaces: {\it i.e.} at the (100) surface (c), there are two Dirac cones at SBZ momentum points  $\overline{\Gamma}$ and $\overline{X}^{\prime}$, at the (001) surface (d) there is no Dirac cone.}
\label{Fig:ws-TB}
\end{figure}

By tuning the parameters $M$, $t_{z}$ and $\lambda$ in Eq.~(\ref{weak TI}), a qualitatively similar electronic structure with a weak topological insulator phase -- like the one of Ca$_{2}$PtO$_{4}$ -- can be obtained.
In Fig.~\ref{Fig:ws-TB} (a), the bulk electronic structure displays two band inversions at $\Gamma$ (0,0,0) and $Z$ (0,0,$\pi$) with $M=1.5$, $t_{z}=0.2$, and $\lambda=0.3$.
These two inversions, when projected onto different surfaces, can give rise to different numbers of TSS, {\it i.e.} a similar situation like what we observed in Ca$_{2}$PtO$_{4}$ (see Fig.~\ref{Fig:SBZ}).
In Fig.~\ref{Fig:ws-TB} (c) and (d), (100) and (001) surfaces are taken as an illustration.
At (100), $\Gamma$ and $Z$ project into $\overline{\Gamma}$ and $\overline{X}^{\prime}$ separately, thus, there are two Dirac cones in the surface Brillouin Zone (SBZ) of the (100) surface.
In contrast, at the (001) surface, there is no TSS, as $\Gamma$ and $Z$ project to the same $\overline{\Gamma}$.

\begin{figure}[htbp]
\centering
\includegraphics[width=0.4\linewidth]{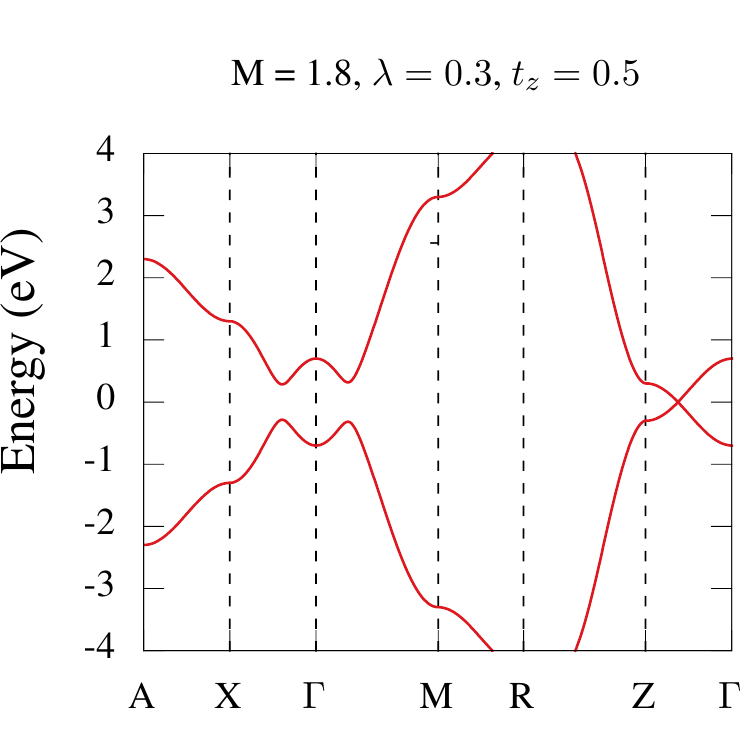}%
\hspace{0.5cm}
\includegraphics[width=0.4\linewidth]{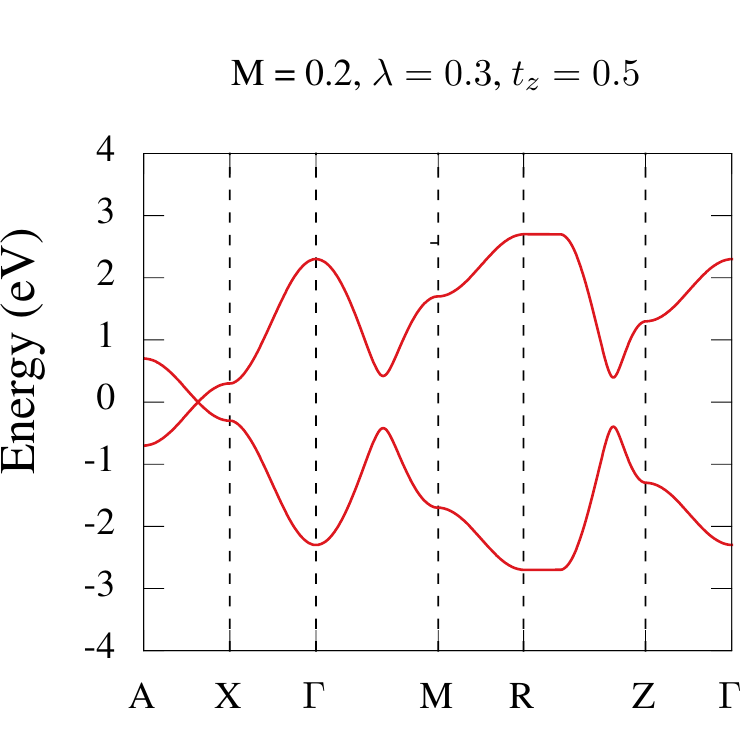}
\caption{(Color online) The bulk band dispersion for the two DSM phases. (left) When $t_{z}<M < (2-t_{z})$, there is a bulk-band linear crossing between $\Gamma$ and $Z$; (right) when $M < t_{z}$, the linear crossing is between $A$ and $X$.}
\label{Fig:DSM}
\end{figure}

In 3D, there are eight TRIMs. This leads to four independent ${\mathbb{Z}}_{2}$ invariants for a gaped system. One
of these invariants, $\nu_{0}$, can be expressed as the product over all eight points,
\begin{equation}\label{Z2-strong}
(-1)^{\nu_{0}}=\prod_{i=1}^{8}\delta_{i}\;,
\end{equation}
where $\delta_{i}$ is the product of the parity eigenvalue for all occupied states at a TRIM $\Gamma_{i}$. In terms of the Hamiltonian~(\ref{TB-Z2}) which has only one degenerate valence band, it take the following simple form~\cite{PhysRevB.76.045302}:
\begin{equation}\label{Z2-define}
\delta_{i}=-\mbox{sgn}(H(\mathbf{k}=\Gamma_{i}))\;.
\end{equation}
The other three invariants are given by products of four $\delta_{i}$'s, for which $\Gamma_{i}$ reside in the same plane,
\begin{equation}\label{Z2-weak}
(-1)^{\nu_{k}}=\prod_{n_{k}=1;n_{j\ne k}=0,1}\delta_{i=(n_{1}n_{2}n_{3})}\;.
\end{equation}
The Hamiltonian (\ref{TB-Z2}) at these momenta reads:
\begin{subequations}
\begin{align}
H(\Gamma) &= [M - (2+t_{z})]\sigma_{z}\otimes I\;; \\
H(2X) &= [M - t_{z}]\sigma_{z}\otimes I\;;\\
H(M) &=[M + 2 - t_{z}]\sigma_{z}\otimes I\;;\\
H(Z) &=[M - 2 + t_{z}]\sigma_{z}\otimes I\;;\\
H(2A) &=[M +  t_{z}]\sigma_{z}\otimes I\;;\\
H(R) &=[M + 2 + t_{z}]\sigma_{z}\otimes I\;.
\end{align}
\end{subequations}
At each TRIM, the Hamiltonian is proportional to the parity operator $\hat{P}=\sigma_{z}\otimes I$, thus, they obviously have the same eigenstates. Furthermore, at each TRIM, two of these states form one occupied Kramers pair and the two other ones correspond to an empty Kramers pair.
If  $M>2+t_{z}$, at all eight TRIMs, the occupied Kramers pair has an eigenvalue of $-1$. Therefore $\delta_{i}=1$ at all eight TRIMs and the topological invariants are simply $0;(000)$, this is the trivial band insulator (BI) phase shown in Fig.~\ref{Fig:TB-topology}.
For $2-t_{z}<M < 2+t_{z}$ there is a bulk-band linear crossing between $Z$ and $\Gamma$ (see the left plot in Fig.~\ref{Fig:DSM}) and the system is a Dirac semimetal. According to the classification for 3D Dirac semimetals, the DSM in our model belongs to the first class that is created via a band inversion~\cite{ncomms5898}, which is topologically nontrivial (see also the topological characterization of this phase in the main text).
The appearance of the bulk-band linear crossing is one striking feature of this 3D topological model, which is resulting from the different inversions at $Z$ and $\Gamma$.
The $k_{z}=0$ and $k_{z}=\pi$ planes are different only at $\Gamma$ and $Z$, as one is inverted and the other one is not.
Thus, when changing $k_{z}=0$ to $k_{z}=\pi$, the topology has to be changed from a nontrivial one to a trivial one.
One way of changing the topology is to close the bulk gap.
Since there is a band inversion at $\Gamma$, this DSM phase is still topologically nontrivial~\cite{ncomms5898}.
The difference to a strong TI is the absence of a finite bulk gap in the DSM. Thus, the corresponding topological surface states in the DSM phase are embedded in the bulk bands.
If $t_{z}<M < (2-t_{z})$, a weak TI phase appears, in which the occupied Kramers pair at both $\Gamma$ and $Z$ have eigenvalues $1$. The corresponding topological invariants are then $0;(001)$.
For $M < t_{z}$ a DSM phase appears again, in which the bands at the two $X$ points are also inverted.
Similarly to the difference between $\Gamma$ and $Z$ for $t_{z}<M < (2-t_{z})$, now, the topology of the $k_{z}=0$ and $k_{z}=\pi$ planes only differ at $X$ and $A$. Thus, by changing $k_{z}$ from $0$ to $\pi$, we have to go through a bulk gap closing between $A$ and $X$ (see the right plot in Fig.~\ref{Fig:DSM}).
The complete phase diagram of the Hamiltonian (\ref{weak TI}) for $M>0$ is displayed in Fig.~\ref{Fig:TB-topology}.
It is symmetric with respect to positive and negative values of $M$.
For negative values of $M$, bands at more TRIMs will be inverted. However, the different phases remain.
Thus, in the following many-body investigations, we only focus on the positive choice of $M$.

When $U\ne0$, we have to solve the Hamiltonian at finite $U$ numerically.
Here, we employ dynamical mean-field theory (DMFT)~\cite{RevModPhys.68.13} with the continuous-time hybridization-expansion (CT-HYB)~\cite{PhysRevLett.97.076405, PhysRevB.74.155107} quantum Monte Carlo as an impurity solver.
The CT-HYB solver is based on the implementation discussed in previous works of us~\cite{PhysRevB.85.115103, PhysRevB.86.155158}.
To illustrate the correlation effect, we set the SOC strength to $\lambda=0.3$ and the inter-layer coupling to $t_{z}=0.5$ throughout the calculations and explore the interacting phase diagram as functions of $M$ and $U$.
The phase boundaries in Fig.~\ref{Fig:PhaseDiagram} are determined from the effective Hamiltonian $H(\mathbf{k})+\Re\hat{\Sigma}(\omega\rightarrow0)$, as explained in the main text. 
The transition lines to the DSM phases are more difficult to be located, as they involve gap closings, which break down the definition of the 3D topological invariants. 
A full analysis would actually require a study of the local spectral function obtained from both the real and the imaginary parts of $\hat{\Sigma}(\omega)$ or a careful examination of the temperature scaling of the Green's function.
Here we have taken a simplified route and looked only at the gap closing induced by the rigid shift $\Re \hat{\Sigma}(\omega\rightarrow0)$, which is accurate mostly in the weak-$U$ region of the phase diagram. 
We note that the precise location of the phase boundaries is not of our prime interest here, as we are interested in the general trend of the phase transformation as a function of $U$.

\begin{figure}[htbp]
\centering
\includegraphics[width=0.4\linewidth]{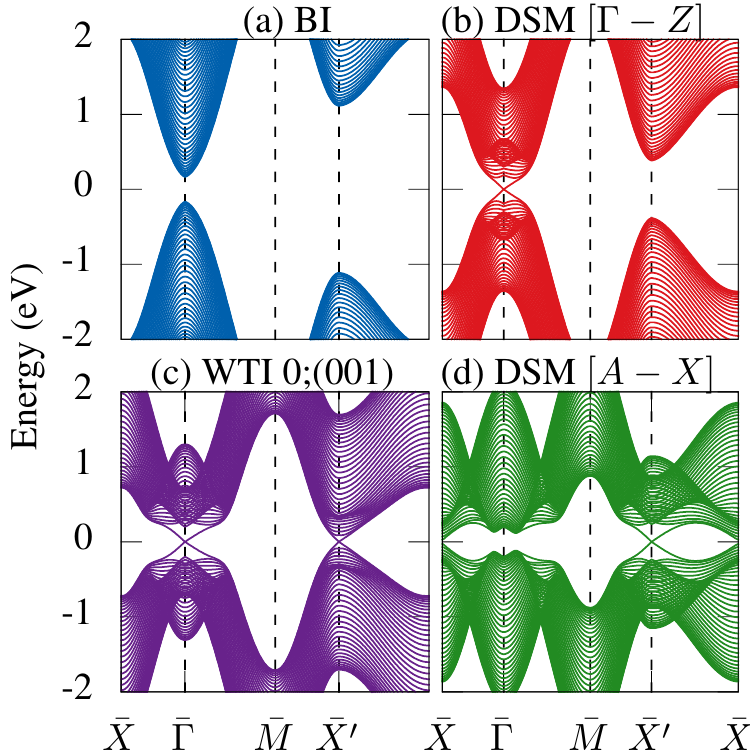} %
\hspace{0.5cm}
\includegraphics[width=0.4\linewidth]{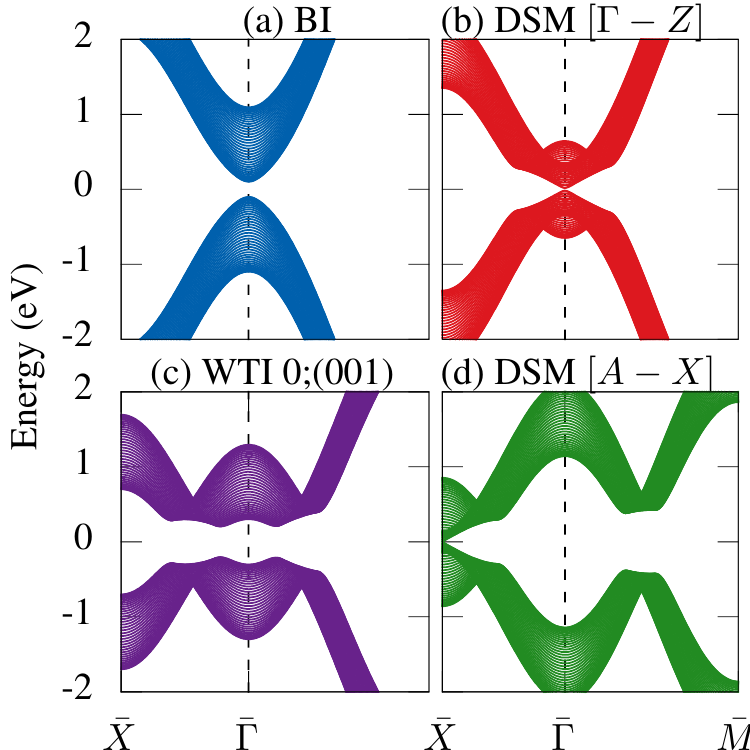}
\caption{(Color online) (left) The topological states at the (100) surface with (a) $U=4.0, M=3.0$; (b) $U=7.0, M=3.0$; (c) $U=8.0, M=3.0$; (d) $U=4.0, M=1.0$. From (a) to (c), the topological phase transitions are driven by the increase of $U$.  (right) Same as the left plot, but for (001) surface. At the two DSM phases, the bulk gaps are closed, thus the TSS at these two phases are embedded in the bulk bands.}
\label{Fig:TB-U}
\end{figure}

In Fig.~\ref{Fig:TB-U}, the topological surface states (TSS) of four topologically distinct phases are displayed with parameters given by the corresponding colored suit symbols shown in Fig.~\ref{Fig:PhaseDiagram}.
The TSS are calculated for the (100) surface (left plot) and (001) surface (right plot) in a 40-layer slab.
For simplicity and to have a better low-energy resolution, we have used the effective one-particle Hamiltonian $h_{\text{eff}}(k)$ with only the real-part of the self-energy $\Re\hat{\Sigma}(\omega\rightarrow0)$ in the calculations.
At $\omega\ne0$, the nonzero value of $\Im\hat{\Sigma}(\omega)$ broadens the electronic bands and smears out the details of the TSS at large frequencies. 
However, the topology of the system is fully determined by $h_{\text{eff}}(\mathbf{k},\omega=0)$ where $\Im\hat{\Sigma}(\omega)$ is zero in all four phases displayed in Fig.~\ref{Fig:TB-U}.
In Fig.~\ref{Fig:TB-U}(a)-(c), we show the TSS with fixed $M=3$ and three different values of $U$.
With the increase of interaction, a topological nontrivial phase can be transformed to a DSM by inverting the bands at the $\Gamma$ point and further to a WTI with invariants $0;(001)$ by also inverting the bands at the $Z$ point. In Fig.~\ref{Fig:TB-U}(d), we show another type of DSM, the corresponding parameters are taken as $M=1, U=4$.
Its TSS also contains two Dirac cones, but their locations in the SBZ are different from that in Fig.~\ref{Fig:TB-U}(c) (see also Fig.~\ref{Fig:TB-topology} for the difference of band inversions).
The left four plots in Fig.~\ref{Fig:TB-U} are the TSS at the (100) surface, which all contain a well-defined band gap, however this is not the case for the (001) surface. At the (001) surface, the two DSM phases are gapless. 
In combination with the left plots, one can easily understand that the TSS in the two DSM phases are embedded in the bulk states.

\end{document}